\documentclass[twocolumn,preprintnumbers,amsmath,amssymb]{revtex4}
\usepackage{dcolumn}
\usepackage{bm}
\usepackage{graphicx}
\usepackage{amsmath}
\usepackage{amsfonts}
\usepackage{amssymb}
\usepackage{amsthm}
\usepackage{amstext}
\usepackage{amsbsy}
\usepackage{amsopn}
\usepackage{amscd}
\usepackage{amsxtra}

\def\phi{\varphi}
\def\epsilon{\varepsilon}

\newcommand{\bec}{\begin{center}}
\newcommand{\enc}{\end{center}}
\newcommand{\be}{\begin{equation}}
\newcommand{\ee}{\end{equation}}
\newcommand{\bmi}{\begin{minipage}}
\newcommand{\emi}{\end{minipage}}

\newcommand{\bi}{\begin{itemize}}
\newcommand{\ei}{\end{itemize}}
\newcommand{\ba}{\begin{array}}
\newcommand{\ea}{\end{array}}

\everymath{\displaystyle}

\begin{document}
\title{Optimal control of an inhomogeneous spin ensemble coupled to a cavity}
\author{Q. Ansel$^{1,2}$}
\author{S. Probst$^4$}
\author{P. Bertet$^4$}
\author{S. J. Glaser$^1$}
\author{D. Sugny$^{2,3}$}
\email{dominique.sugny@u-bourgogne.fr}

\affiliation{$^1$ Department of Chemistry, Technische Universit\"at
M\"unchen, Lichtenbergstrasse 4, D-85747 Garching, Germany}
\affiliation{$^2$ Laboratoire Interdisciplinaire Carnot de Bourgogne (ICB), UMR 6303 CNRS-Universit\'e de Bourgogne- Franche Comt\'e, 9 Av. A. Savary, BP 47 870, F-21078 DIJON Cedex, FRANCE}
\affiliation{$^3$ Institute for Advanced Study, Technische Universit\"at M\"unchen, Lichtenbergstrasse 2 a, D-85748 Garching, Germany}
\affiliation{$^4$ Quantronics group, Service de Physique de l'Etat Condense, DSM/IRAMIS/SPEC,CNRS UMR 3680, CEA-Saclay, 91191 Gif-sur-Yvette cedex, France}
\date{\today}
\begin{abstract}
We apply optimal control techniques to an inhomogeneous spin ensemble coupled to a cavity. A general procedure is proposed for designing the control strategies. We numerically show the extent to which optimal control fields robust against system uncertainties help enhancing the sensitivity of the detection process. The parameters of the numerical simulations are taken from recent Electron Spin Resonance experiments. The low and high cooperativity regimes are explored.
\end{abstract}

\maketitle

\section{Introduction}
Quantum optimal control theory (QOCT) is aimed at finding a way to bring a quantum system from one state to another with minimum expenditure of time and resources \cite{glaserreview,brifreview,dongreview,altafinireview,alessandrobook,ricebook,bonnardbook}. Intense progress has been realized recently in the development of such techniques \cite{glaserreview,brifreview}. Several optimization algorithms \cite{grape,reichkrotov,gross,calarco} have been proposed to design control fields suited to different experimental setups and constraints or robust against experimental uncertainties and modelling imperfections \cite{kobzar1,kobzar2,kobzar3,rabitzrobust,rabitzrobust2,khanejaens,li1,li2,daemsprl}. QOCT has been first developed in molecular physics to steer chemical reactions \cite{brifreview,ricebook,somloi,zhu} and in Nuclear Magnetic Resonance (NMR) or Magnetic Resonance Imaging for controlling spin dynamics \cite{grape,lapertprl,contrast,vanreeeth:2017_1,vanreeth:2017,lefebvre:2017,conolly,maximov:2010,maximov:2015}. OCT is nowadays attracting a lof of effort in the context of quantum information processing \cite{goerz,garon,khanejaspin,kontz} and has been recognized as one of the cornerstones for enabling quantum technologies \cite{glaserreview}.

In NMR, spin dynamics are governed by the Bloch equation and controlled by a radio-frequency magnetic field which is assumed to be a piece-wise constant function adjustable in time. This approximation corresponds to a standard framework in QOCT~\cite{grape}. The situation is not so simple in Electron Spin Resonance (ESR) and specific experimental constraints due to technical limitations of the spectrometer have to be accounted for. For instance, the role of the microwave resonator cannot be neglected and the field applied to the spins is distorted by the response function of the cavity~\cite{bookEPR,spindler:2012}. The main experimental limitation is the nonlinearity of the resonator which arises for large amplitudes of the intra-cavity field, particularly for superconducting micro-resonators. In other experimental setups, a continuous variation of the amplitude and phase of the control pulses is not possible and only switches between a discrete set of pulse phases is permitted by the available hardware~\cite{discretebook1,discretebook2}. QOCT has been applied with success in these different settings~\cite{spindler:2012,dridi:2015}, showing the efficiency and the flexibility of this approach.

\begin{figure}[h]
	\includegraphics[width=4.5cm]{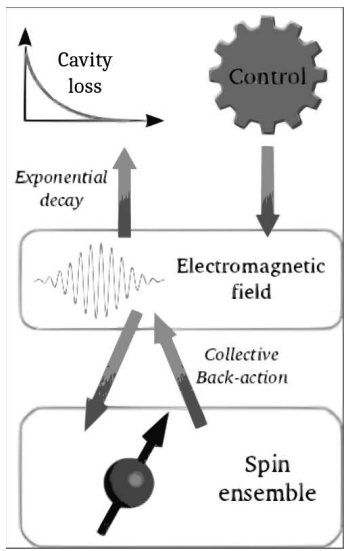}
	\caption{Schematic description of the system (\ref{eqsemi}). The control pulses can only change the electromagnetic field of the cavity and not directly the dynamics of the spins. Note the back-action of the spins onto the cavity mode.}
	\label{fig1}
\end{figure}

The detection of individual spins is a challenging issue in magnetic resonance~\cite{wrachtrup:2016}. Different experimental strategies have been proposed up to date to reach this single spin limit~\cite{wrachtrup:1993,baumann:2015,rugar:1992,gruber:1997,rugar:2004}. Among other propositions, a promising option is to push to its physical limit the inductive detection method in ESR~\cite{molmer:2013,bienfait:2015,bienfait:2016,bienfait:2017,prost:2017}. Recent progress has shown that 260 spins per echo can by now be detected with signal-to-noise of 1~\cite{prost:2017}. This gain of several orders of magnitude in sensitivity over the conventional approaches has been made possible by different experimental advances extending from the cryogenic temperature of the sample to the high quality factor of the micro-resonators and by the use of Josephson parametric amplifier devices. Hahn echo or CPMG sequences~\cite{levitt_book} are usually implemented with standard rectangular pulses to measure the echo signal emitted by the spins in the cavity~\cite{levitt_book,ernst_book,hahn_paper}. However, the efficiency of these control protocols is limited by the response function of the resonator and by the inhomogeneities and imperfections of the sample. In the running to the single-spin detection, it is therefore crucial to identify control procedures enhancing the echo signal for a given number of excited spins. This issue is addressed numerically in this paper by using tools of OCT, which offer the possibility to go beyond intuitive protocols. For sake of concreteness, the theoretical analysis of this work is based on some recent experiments made in ESR~\cite{bienfait:2015,bienfait:2016,bienfait:2017,prost:2017}.
A schematic description of the physical concepts at the basis of the control process in given in Fig.~\ref{fig1}. The control of the spin ensemble is a two-step procedure in which only the intra-cavity field can be directly modified by the external control, the field applied to the spins being distorted by the response function of the resonator. In the case the cavity acts as a linear bandpass filter, the cavity response can be deconvoluted and the intra-cavity field can be designed (up to some extent) for any given field the spins are subjected to. The back-action of the spin ensemble to the cavity adds, in the high cooperativity regime, a degree of complexity to this control scenario.

From a theoretical point of view, the quantum dynamics is governed by a damped Jaynes-Cummings model. The optimization procedure presented in this paper is an extension of a standard iterative algorithm, namely GRAPE~\cite{grape}. We first optimize the field acting on the spin ensemble to realize efficient Hahn and CPMG sequences even if the system parameters are known with a finite precision. The deconvolution of the resonator response leads then to the intra-cavity field. Specific constraints are accounted for in the optimization process to design realistic fields. We show the efficiency of the corresponding optimal fields for enhancing the signal to noise ratio (SNR) of the detection process and its sensitivity, that is the minimum number of spins per echo that can be detected with signal-to-noise ratio of 1. Note that closely related works have recently investigated the optimal control of such systems for quantum information applications~(see e.g \cite{rojan:2014,allen:2017,krimer:2017,deffner:2014,fisher:2010,sun:2016,freitas:2017} to cite a few).

The remainder of this paper is organized as follows. The model system is introduced in Sec.~\ref{sec2} with special attention paid to the different approximations. Optimal control techniques and robust control fields are described in Sec.~\ref{sec3}. Section~\ref{sec4} presents numerical results based on recent ESR experiments. We conclude in Sec.~\ref{sec6} with an outlook and different perspectives. Supplementary data are reported in Appendix~\ref{appa}.
\section{Theoretical description}\label{sec2}
\subsection{The model system}
We consider an inhomogeneous ensemble of spin 1/2 particles with different resonance frequencies coupled to a single-mode cavity. The parameter values are chosen to be in accordance with recent experiments in ESR~\cite{bienfait:2015,bienfait:2016,bienfait:2017,prost:2017}. The dynamics of the system can be described by the damped Jaynes-Cummings model. In a frame rotating at $\omega$, the frequency of the microwave drive, the density matrix $\rho$ satisfies the following differential equation (in units of $\hbar$):
\begin{equation}\label{eq1}
\dot{\rho}=-i[H,\rho]+\mathcal{L}(\rho),
\end{equation}
where
$$
H=\omega_0 a^\dagger a+\sum_{j=1}^N[\frac{\omega_j}{2}\sigma_z^{(j)}+g_j(a^\dagger \sigma_-^{(j)}+a\sigma_+^{(j)})]+i(\beta a^\dagger -\beta^*a),
$$
and
$$
\mathcal{L}(\rho)=\kappa (a\rho a^\dagger-\frac{1}{2}\rho a^\dagger a-\frac{1}{2}a^\dagger a\rho).
$$
The parameters $\omega_0$ and $\omega_j$ are respectively the detunings of the cavity and of the spins with respect to the drive frequency of the field. We denote by $N$ the number of spins in the ensemble. A specific example will be investigated in Sec.~\ref{sec4}. The coupling strength between each spin and the cavity is given by the constant $g_j$. The amplitude of the microwave drive applied to the cavity mode is represented by the time-dependent functions $\beta(t)$ and $\beta^*(t)$. The cavity losses are described by the Lindbladian $\mathcal{L}$ with a rate $\kappa$. We recall the standard commutation relations between the different operators: $[a,a^\dagger]=1$, $[\sigma_x,\sigma_y]=2i\sigma_z$, $[\sigma_+,\sigma_-]=\sigma_z$, $[\sigma_z,\sigma_\pm]=\pm 2\sigma_\pm $, $\sigma_x=\sigma_++\sigma_-$ and $\sigma_y=-i(\sigma_+-\sigma_-)$. Using Eq.~\eqref{eq1}, it is straightforward to show that the time evolution of the expectation values of the different operators, denoted $\langle\cdot\rangle$, is governed by:
\begin{equation}\label{eqini}
\begin{cases}
\frac{d}{dt}\langle \hat{X}\rangle =-\omega_0\langle \hat{Y}\rangle-\frac{\kappa}{2}\langle \hat{X}\rangle +\omega_X -\sum_j2g_j\langle \hat{S}_y^{(j)}\rangle \\
\frac{d}{dt}\langle \hat{Y}\rangle =\omega_0\langle \hat{X}\rangle-\frac{\kappa}{2}\langle \hat{Y}\rangle+\omega_Y +\sum_j 2g_j\langle \hat{S}_x^{(j)}\rangle \\
\frac{d}{dt}\langle \hat{S}_x^{(j)}\rangle =-\omega_j \langle \hat{S}_y^{(j)}\rangle+g_j\langle \hat{Y}\hat{S}_z^{(j)}\rangle \\
\frac{d}{dt}\langle \hat{S}_y^{(j)}\rangle =\omega_j \langle \hat{S}_x^{(j)}\rangle-g_j\langle \hat{X}\hat{S}_z^{(j)}\rangle \\
\frac{d}{dt}\langle \hat{S}_z^{(j)}\rangle =g_j\langle \hat{X}\hat{S}_y^{(j)}\rangle-g_j\langle \hat{Y}\hat{S}_x^{(j)}\rangle,
\end{cases}
\end{equation}
where we have introduced the operators $\hat{S}_{x,y,z,\pm}^{(j)}=\sigma_{x,y,z,\pm}^{(j)}/2$, $\hat{X}=a+a^\dagger$ and $\hat{Y}=-i(a^\dagger-a)$ in order to get more symmetrical expressions. The control fields are given by the two real functions $\omega_X=\beta+\beta^*$ and $\omega_Y=i(\beta-\beta^*)$.

From now on, we consider different degrees of approximation which lead to numerical optimizations computationally less expensive. These approximations are widely used in quantum optics~\cite{haroche_book}. A first approximation consists in neglecting quantum correlations between the spins and the cavity mode. This approximation can be introduced in the framework of cumulant expansion~\cite{JC,kubo}. In this work, we consider an expansion of order two in which $\langle \hat A \hat B \rangle \simeq \langle \hat A \rangle \langle \hat B \rangle$. This approximation is justified in the bad cavity limit and it greatly simplifies the differential system which becomes of dimension $3N+2$. In the frame rotating at frequency $\omega_0$, we have:
\begin{equation}\label{eqsemi}
\begin{cases}
\dot{X} =-\frac{\kappa}{2}X +\omega_X -\sum_j2g_j S_y^{(j)} \\
\dot{Y} =-\frac{\kappa}{2}Y+\omega_Y +\sum_j 2g_jS_x^{(j)} \\
\dot{S}_x^{(j)} =-\Delta_j S_y^{(j)}+g_jYS_z^{(j)} \\
\dot{S}_y^{(j)} =\Delta_j S_x^{(j)}-g_j XS_z^{(j)} \\
\dot{S}_z^{(j)} =g_jXS_y^{(j)}-g_jYS_x^{(j)},
\end{cases}
\end{equation}
where we use (for sake of simplicity) the same notations $A=\langle\hat{A}\rangle$ for the observable $A$ in the initial and the rotating frames. For each spin, we introduce the offset $\Delta_j = \omega_j-\omega_0$. In Eq.~\eqref{eqsemi}, $X$, $Y$, $\omega_X$, $\omega_Y$ and $\vec{S}^{(j)}$ are expressed in dimensionless units. A straightforward physical interpretation of the dynamics can be given from Eq.~\eqref{eqsemi}. Equation~\eqref{eqsemi} describes an ensemble of spins coupled to a driven classical oscillator, whose dynamics depends on the state of the spins. In particular, we obtain that the spins move along the Bloch sphere, $\vec{S}^{(j)}(t)^2=\textrm{cst}$. This description is illustrated in Fig.~\ref{fig1}. This situation is similar to the one encountered in Magnetic Resonance when the detection coil induces a back action on the spins. This well-known effect, called the radiation damping effect, is modeled by non-linear terms in the Bloch equation governing the spin dynamics~\cite{zhang:2011,RD1,RD2,RD3}. This modelling can be recovered from Eq.~\eqref{eqsemi} in the bad cavity limit discussed in Sec.~\ref{sec:bad cavity limit} where $g\ll \kappa$~\cite{azouit1,azouit2,julsgaard:2012,wang:2005,atkins:2003,brion:2007}.

The spin ensemble is also subjected to standard $T_1$ and $T_2$ relaxation processes. The interaction of the spins with the cavity leads to another $T_1-$relaxation process, called the Purcell effect \cite{haroche_book,purcell_th,bienfait:2016}, which can be interpreted as a cavity-enhanced spontaneous emission. The corresponding relaxation time $T_1^p$ can be expressed as:
\begin{equation}\label{eqpurcell}
\frac{1}{T_1^p}=\frac{\kappa g_i^2}{\Delta_j^2+\kappa^2/4}.
\end{equation}
Note that spins detuned from the cavity have a longer relaxation time and that $T_1^p\ll T_1$ in the experimental situation we consider in this work.

Finally, we point out that some collective effects of the spin ensemble, such as super-radiant relaxation, governed by the Tavis-Cummings Hamiltonian~\cite{tavis:1968} occur if the offset inhomogeneities are not too strong~\cite{butler:2011,temnov:2005,wood:2016}. The transition between the individual and collective spin regimes is described by the cooperativity parameter $C$. In the case of a spin ensemble with a Lorentzian density of frequencies of full width at half height $\Omega$, the parameter $C$ can be expressed as $C=\frac{2Ng^2}{\kappa \Omega}$~\cite{temnov:2005}.

\subsection{The bad cavity limit}
\label{sec:bad cavity limit}

We recall in this section the standard approximation of the bad cavity limit. We first integrate the two first equations of \eqref{eqsemi} and we obtain:
\begin{equation}
\begin{cases}
X(t) =\int_{-\infty}^t e^{-\frac{\kappa}{2}(t-t')}C_X(t')dt' \\
Y(t) =\int_{-\infty}^t e^{-\frac{\kappa}{2}(t-t')}C_Y(t')dt',
\end{cases}
\end{equation}
with $C_X(t') = \omega_X(t') -2\sum_jg_jS_y^{(j)}(t')$ and $C_y(t') = \omega_Y(t') + 2\sum_jg_jS_X^{(j)}(t')$.
We define the integral $\mathsf{F}$ of a function $f$:
$$
\mathsf{F}=\int_{-\infty}^t e^{\frac{\kappa}{2}(t'-t)}f(t')dt',
$$
the goal being to compute the limit of $\mathsf{F}$ when $\kappa\to +\infty$. For that purpose, we define a sequence of functions $\mu_k$
$$
\mu_k(t)=\frac{\kappa_k}{4}e^{-\kappa_k|t|/2},
$$
which converges towards the Dirac distribution if $\kappa_k\to +\infty$ when $k\to +\infty$. From the relations of generalized function theory which involve the product of a Dirac distribution with a Heaviside function~\cite{collombeau}, we obtain that:
$$
\mathsf{F}\simeq \frac{2}{\kappa}f(t)
$$
when $\kappa\gg 1$ ($\kappa$ is expressed here in the dimensionless units of Eq.~\eqref{eqsemi}). This approximation leads to:
\begin{equation}\label{eqxy}
\begin{cases}
X(t) = \frac{2}{\kappa}(\omega_X-2\sum_jg_jS_y^{(j)})\\
Y(t) = \frac{2}{\kappa}(\omega_Y+2\sum_jg_jS_x^{(j)}).
\end{cases}
\end{equation}
Plugging \eqref{eqxy} into \eqref{eqsemi}, we arrive at:
\begin{equation}\label{rd1}
\begin{cases}
\dot{S}_x^{(j)} =-\Delta_j S_y^{(j)}+\frac{2g}{\kappa}\omega_YS_z^{(j)}+\frac{4g^2}{\kappa}\bar{S}_x S_z^{(j)} \\
\dot{S}_y^{(j)} =\Delta_j S_x^{(j)}-\frac{2g}{\kappa}\omega_XS_z^{(j)}+\frac{4g^2}{\kappa}\bar{S}_y S_z^{(j)} \\
\dot{S}_z^{(j)} =\frac{2g}{\kappa}(\omega_XS_y^{(j)}-\omega_YS_x{(j)})-\frac{4g^2}{\kappa}(\bar{S}_x S_x^{(j)}+\bar{S}_y S_y^{(j)}),
\end{cases}
\end{equation}
where $g_j=g$ is assumed to be the same for all the spins and $\bar{S}_x=\sum_k S_x^{(k)}$, $\bar{S}_y=\sum_k S_y^{(k)}$. We recover the standard equations describing the radiation damping effect with a rate of $\frac{4g^2}{\kappa}\bar{S}_{x,y}$. Note here the unusual sign convention of this term since the thermal equilibrium point of the Bloch ball is the south pole and not the north pole of the Bloch sphere. This analysis also highlights a difference between the radiation damping and the Purcell effect since this latter occurs at rate $\frac{4g^2}{\kappa}$ (see Eq.~\eqref{eqpurcell} for $\Delta_j=0$) even when $\bar{S}_{x,y}=0$.  Finally, note that the bad cavity limit leads to a standard optimal control problem in magnetic resonance~\cite{glaserreview,nielsen:2010}, which was solved e.g. in~\cite{zhang:2011}.
\section{Optimal control of a spin ensemble}\label{sec3}
\subsection{The general procedure}
We propose in this paragraph a general procedure to design optimal control fields in a dynamical system governed by Eq.~\eqref{eqsemi}. The inspection of the system~\eqref{eqsemi} clearly shows that the control fields $\omega_X$ and $\omega_Y$ only change the time evolution of the quadratures $X$ and $Y$ and not directly the spin dynamics, which can only be modified in a two-step process. The two control fields can be expressed in terms of $X$ and $Y$ through the response function of the cavity:
\begin{equation}\label{eq:control}
\begin{cases}
\omega_X = \dot{X} + \frac{\kappa}{2}X +\sum_j2g_j S_y^{(j)}  \\
\omega_Y = \dot{Y} + \frac{\kappa}{2}Y - \sum_j 2g_jS_x^{(j)}. \\
\end{cases}
\end{equation}
For fast changing cavity fields, we observe in Eq.~\eqref{eq:control} that the time derivative of $X(t)$ and $Y(t)$ diverges and so does the control amplitudes $\omega_X$ and $\omega_Y$.

The design process of the control fields can be decomposed into three different steps. We first determine the time evolution of $X(t)$ and $Y(t)$ to realize a given control task on the spin system. Note that this control issue is a standard control problem of a spin ensemble with offset (the parameters $\Delta_j$) and field (the parameters $g_j$) inhomogeneities (see Ref.~\cite{glaserreview,kobzar1,kobzar2,kobzar3} for a series of results on the subject). We compute numerically, in a second stage, $\sum_jg_jS_x^{(j)}$ and $\sum_jg_jS_y^{(j)}$ with the last three equations of \eqref{eqsemi}. This step can be neglected in the low coupling limit ($g_j \simeq 0$). The fields $\omega_X$ and $\omega_Y$ are finally obtained with Eq.~\eqref{eq:control}. The only mathematical constraint of this procedure relies on the fact that $X(t)$ and $Y(t)$ must be differentiable functions. This assumption will be satisfied by expanding $X$ and $Y$ on a specific function basis. Experimental limitations on the control fields will be discussed in Sec.~\ref{sec4}.

Another option for the optimization process is to consider the dynamical system as a whole and to define a control objective in terms of the spin and the cavity coordinates. This approach was not used in this work for two main reasons. Due to the non linear character of the dynamics, it is no more possible to define straightforwardly universal rotations pulses~\cite{kobzar2}, i.e. rotations which do not depend on the initial state of the spin. Such rotations are an essential building block of spin echo or CPMG sequences~\cite{levitt_book,coryCPMG:2010}. In addition, we have observed that the corresponding control landscape admits many local extrema which prevent a fast convergence of the algorithm.
We show below on different benchmark examples how to control the spin ensemble. For the sake of generality, all the parameters are expressed in dimensionless units in Sec.~\ref{sec3}. A dimensional analysis can be used to determine the physical units of the different coefficients. If $t_f$ denotes the real control time then $t\mapsto t\times t_f$, $\kappa\mapsto \kappa/t_f$, $\omega_{X,Y}\mapsto \omega_{X,Y}/t_f$, $g\mapsto g/t_f$ and $\Delta_j\mapsto \Delta_j/t_f$. Note that the $X$, $Y$ and the $\vec{S}$- variables remain dimensionless.
\subsection{Bump pulses}

\begin{figure}[h]
\includegraphics[width=8cm]{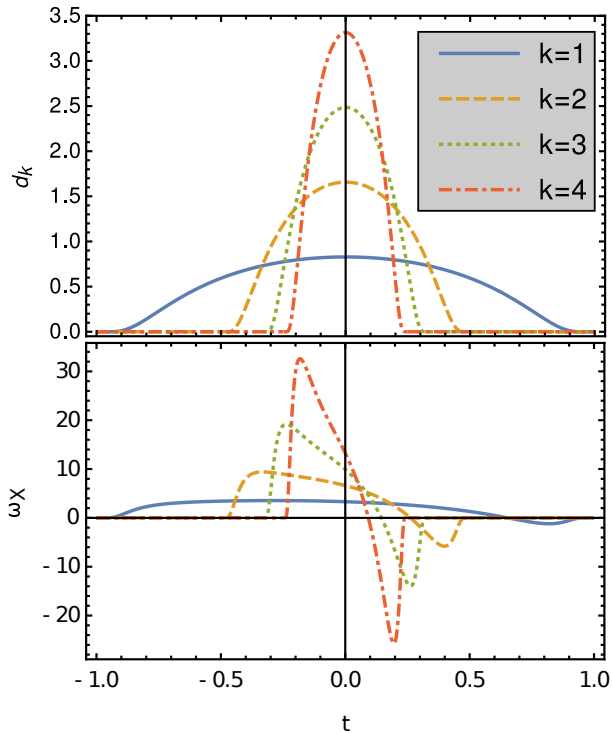}
\caption{(Color online) (top) Sequence of functions $d_k(t)$ with a compact support converging to the Dirac distribution for $k=1$, $2$, $3$ and $4$. (bottom) Plot of the corresponding control field $\omega_X$ with $\kappa=4$ (the back action from the spin system is neglected). Dimensionless units are used.}
\label{fig2}
\end{figure}

The first control sequence is aimed at reproducing the effect of a Dirac pulse on a spin system~\cite{levitt_book,ernst_book}. Dirac pulses are generally approximated by square pulses with an area corresponding to the rotation angle induced by the field. Here, square pulses are not suited to the control process because their derivatives are singular distributions. Instead we propose to use the space of infinitely smooth bump functions of compact support~\cite{schwartzbook}, as displayed in Fig.~\ref{fig2}. They have the property to be infinitely differentiable and to go to zero on the boundaries of their support. We introduce the set of functions $(d_k)_{k\in \mathbb{N}}$ defined by:
\begin{equation}
d_k(t) =A ke^{1/(k^2t^2-1)}\mathbb{I}_{[-1/k,1/k]}(t),
\end{equation}
where $\mathbb{I}_I$ is the indicator function on the interval $I$ (this function takes the value 1 for elements of $I$ and 0 outside). The bump functions satisfy:
$$
\lim_{k \rightarrow \infty} d_k = \delta.
$$
The pulse duration $t_f$ is set by the parameter $k$, $t_f = 2/k$. The normalization factor $A$ is chosen so that $\int_{-1/k}^{1/k} d_k(t) dt = 1$. It can be expressed in terms of Whittaker's function: $ A= \sqrt{\pi/e}W_{-1/2,1/2}(1) = 0.44399\cdots $.
An approximate Dirac pulse is applied to the spin system in the $x$- direction if $X(t)=\theta d_k(t)$, where $\theta$ is the rotation angle induced by the pulse. A specific value of $k$ is chosen. Using Eq.~\eqref{eqxy}, this leads to the following control field $\omega_X$:
\begin{equation}
\begin{split}
\omega_X(t) = & \theta A k\left( \kappa-\frac{2k^2t}{(k^2t^2-1)^2}\right)e^{1/(k^2t^2-1)} \\
&\times \mathbb{I}_{[-1/k,1/k]}(t) + \sum_j2g_j S_y^{(j)}.
\end{split}
\label{eq:bump_pulse}
\end{equation}
Figure~\ref{fig2} shows an example of different pulse sequences. This family of pulses is called hereafter \emph{bump pulses}.

\subsection{Design of optimal control fields}

\begin{figure}[h]
\centering
\includegraphics[width=8.5cm]{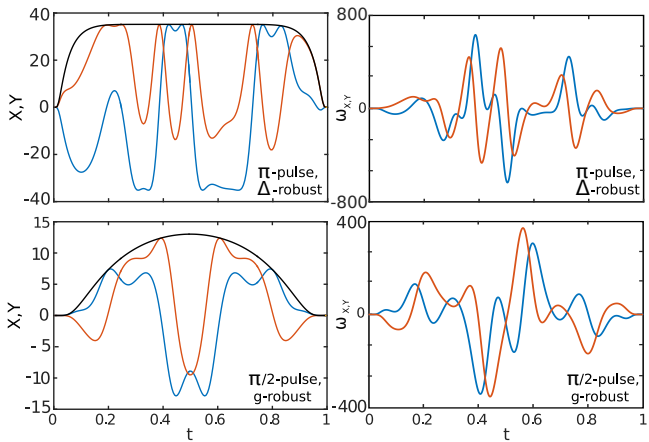}
\caption{(Color online) \textit{left} - Optimized $\pi$-  (top) and $\pi/2$- pulses (bottom) robust against offset or coupling strength distribution. $X$, $Y$ and $(X^2+Y^2)^{1/2}$ are displayed respectively in red (dark gray), blue (black) and black solid lines. \textit{right} - Plot of the corresponding control fields $\omega_X$ and $\omega_Y$. Dimensionless units are used. The parameter $\kappa$ is set to 4.}
\label{fig3}
\end{figure}

More general control protocols can be designed along the same method. We consider the excitation and the inversion of the spin ensemble with different offset and field inhomogeneities. These robust pulse sequences are nowadays standard in magnetic resonance~\cite{glaserreview,nielsen:2010,kobzar1,kobzar2,kobzar3,coryCPMG:2010} either for point to point transformations or universal rotations. We investigate below this second class of pulses. The optimization procedure requires the introduction of a figure of merit $\mathcal{F}$ to maximize at time $t=t_f$. For the stability of numerical optimization, we come back to the Schr\"odinger picture and matrices of $SU(2)$ are used to describe the rotation of each spin~\cite{kobzar2}. In this setting, $\mathcal{F}$ can be expressed as follows:
\begin{equation}
\mathcal{F}= \frac{1}{2N}\sum_{j=1}^N\Re \left(\text{Tr}\left[U_{f}^\dagger U_j(t_f) \right]\right),
\end{equation}
where $U_f$ is the target evolution operator for each spin and $U_j(t_f)$ the propagator at time $t=t_f$ of the spin $j$. The procedure is illustrated by two specific examples, namely $\pi/2$- and $\pi$- rotations around the $y$- axis, which correspond to the following unitary matrices:
\begin{equation}
U_{\pi} = \left(
\begin{array}{cc}
0 & -1 \\
1 & 0
\end{array}
\right) ~;~
U_{\pi/2} =\frac{1}{\sqrt{2}} \left(
\begin{array}{cc}
1 & -1 \\
1 & 1
\end{array}
\right).
\end{equation}
A peculiar analytical expression for the quadratures $X$ and $Y$ is chosen to design experimentally relevant control fields. They are parameterized as follows:
\begin{equation}\label{eq:opt_X}
\begin{cases}
X = A(t)\cos(\phi(t)) \\
Y = A(t)\sin(\phi(t))
\end{cases}
\end{equation}
with
\begin{equation}\label{eq:phase_opt_control}
\begin{cases}
 A(t)     &= A_0\exp\left(\frac{1}{(2t-1)^p-1}\right)\mathbb{I}_{[0,1]}(t), \\
 \phi(t)  &= \frac{a_0}{2}+\sum_{n=1}^{N_F} a_n\cos(2\pi nt)+ b_n\sin(2\pi nt)
\end{cases}
\end{equation}
where $A_0$ is a normalization factor setting the pulse energy, $p$ is an arbitrary odd number and $\{a_n,b_n\}_{n=0..N_F+1}$ is the set of $2(N_F+1)$ parameters to optimize.
Numerical computations have been performed for two different situations. We first consider a uniform offset distribution in the interval $\Delta_j\in [-30,30]$ and a constant coupling strength, $g=g_0=1$.  In a second example, we assume that all the spins have the same frequency and that the parameter $g$ is given by $g=g_0(1+\alpha)$ with $\alpha\in [-0.3,0.3]$. In the two cases, the control time is set to 1.
Different pulse shapes are plotted in Fig.~\ref{fig3}. The values of the pulse parameters are gathered in App.~\ref{appa}. Note the more than 10 times larger amplitudes obtained for the pulses $\omega_X$ and $\omega_Y$ with respect to the ones of bump pulses in Fig.~\ref{fig2}. A robust pulse against both offset and coupling strength inhomogeneities can also be designed along the same lines but at the price of a longer control time or a larger pulse energy.
%

\section{Numerical results}\label{sec4}
\subsection{Low cooperativity regime}

We investigate an example in the low cooperativity regime reproducing recent experiments in ESR~\cite{bienfait:2015,bienfait:2016,bienfait:2017,prost:2017}. The measured signal in the cavity is an echo signal in $X$ and $Y$ produced by a standard Hahn sequence~\cite{levitt_book,hahn_paper} of the form $(\pi/2)_y -\tau - (\pi)_y - \tau$, where $\tau$ is the echo time. This sequence is repeated several times at a rate $\gamma_r=10~\textrm{Hz}$. In these experiments, the offset distribution ($\simeq~\textrm{MHz}$) is wider than the cavity bandwidth ($\simeq~\textrm{kHz}$). The spin polarization due to the repetition of the experiment reduces of several orders of magnitude the offset bandwidth contributing to the signal. Simulations are thus performed with an effective spin distribution which depends on the polarization $p=1-\exp[-1/(\gamma_r T_1^p)]$. The relaxation times are taken to be $T_1=3$~s, $T_2=1.7$~ms and $T_1^p\simeq 100$~ms for $\Delta=0$. Here, the Purcell rate provides the time required for a spin to relax toward its ground state. The spin does not reach the thermal equilibrium point between two repetitions. We assume that the initial distribution is approximately uniform on the interval $\pm 1.9~\textrm{MHz}$ with a total number of spins $N = 13500$. We also set $\kappa = 9.8\times 10^5~\textrm{s}^{-1}$ and $g_0/(2\pi)=424~\textrm{Hz}$. With such parameters, the effective spin distribution is approximately Lorentzian of full width at half maximum (FWHM) $\Omega_p = 159.15$~kHz. The effective number of spins $N_{\textrm{eff}}$, i.e. the maximum number of spins which can be excited, is defined by:
\begin{equation}
N_{\textrm{eff}}=N\int_{-\infty}^{+\infty}p(\Delta)d\Delta\simeq 940 .
\end{equation}
$N_{\textrm{eff}}$ can also be expressed in terms of the spin components as $N_{eff}=\bar{S}_z(t=0^-)$ where $t=0$ is the time at which the $\pi/2$- pulse of the echo sequence is applied. In this example, the cooperativity parameter $C=\frac{2 N_{\textrm{eff}}g_0^2}{\kappa \Omega_p}$ is of the order of 0.01.

Following Ref.~\cite{bienfait:2015,bienfait:2016,bienfait:2017,prost:2017}, we finally define the number of spins $N_{\textrm{spins}}$ which are effectively excited by the control sequence applied in the interval $[0,t_f]$ and which therefore contribute to the echo signal:
\begin{equation}
N_{\textrm{spins}}=\bar{S}_z(t=0^-)-\bar{S}_z(t_f).
\end{equation}
For a perfectly selective $\pi/2$- pulse of bandwidth $\Omega_c$, $N_{\textrm{spins}}$ can also be estimated as:
\begin{equation}
N_{\textrm{spins}}=N\int_{-\Omega_c/2}^{+\Omega_c/2}p(\Delta)d\Delta .
\end{equation}
In order to improve the sensitivity of the experiment, we are interested in maximizing the signal to noise ratio (SNR) associated with the echo signal in the cavity. The SNR for a single echo can be defined as follows~\cite{bienfait:2015}:
\begin{equation}
\textrm{SNR} = \sqrt{\frac{\int_{echo}  X^2(t)dt}{\Delta X^2}},
\end{equation}
where only the echo along the $X$- direction is accounted for. At a temperature close to 0~K, the noise $\Delta X$ is estimated to be of the order of 1/2, which corresponds to the electromagnetic quantum fluctuations in the cavity, all the other noise sources are neglected~\cite{bienfait:2015,bienfait:2016,bienfait:2017,prost:2017}. The number of excited spins $N_{\textrm{min}}$ for a SNR of 1 is given by $N_{\textrm{min}}=N_{\textrm{spins}}/\textrm{SNR}$ and will be used with the SNR to estimate the efficiency of the excitation process and the sensitivity of the detection.

\begin{figure}[h]
\includegraphics[width=8.5cm]{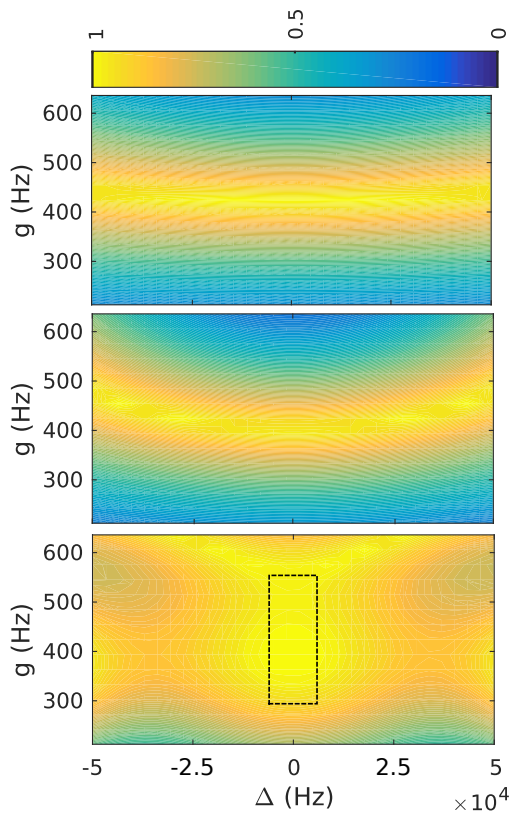}
\caption{(Color online) Robustness of a $\pi/2$- excitation process against coupling strength $g$ and offset $\Delta$ inhomogeneities of a bump pulse (top), a square pulse (middle) and a $g$- robust pulse (bottom). The rectangle in dashed lines indicates the spins used in the optimization. In order to provide a fair comparison, we fix the maximum value of $X$ during the pulse. The pulse duration is set to satisfy this constraint: $t_{\textrm{square}} = 1 ~\mu$s, $t_{\textrm{bump}} = 2t_k=3.9 ~\mu$s and $t_{\textrm{g-robust}} = 19.5 ~\mu$s.}
\label{figrobust}
\end{figure}

\begin{figure}[h]
\centering
\includegraphics[width=8.5cm]{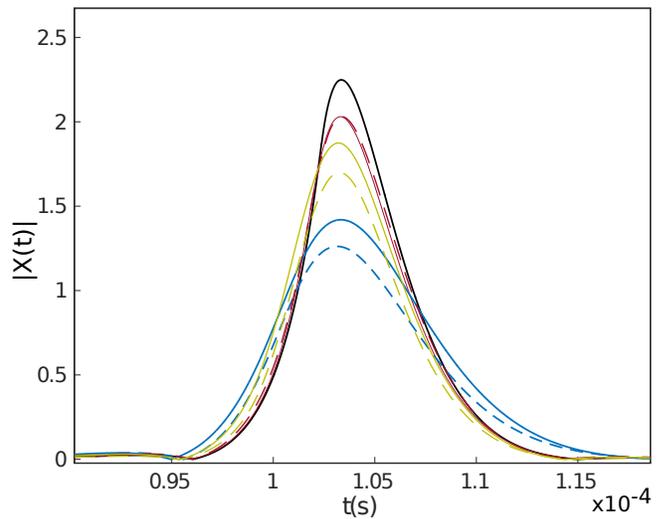}
\caption{(Color online) Comparison of the echo signal in $X$ for square pulses (in blue), for bump pulses (in green) and for $g$- robust pulses (in red). The physical limit with ideal spin rotations is displayed in black. The solid and dashed lines depict respectively the echo signal without and with $g$- inhomogeneities. The pulse duration is set to $1~\mu$s.}
\label{figecho}
\end{figure}

We first investigate the robustness of the excitation process against offset and coupling strength inhomogeneities for the bump, square and $g$- robust pulses. The $g$- robust control sequence is determined by computing $X$ and $Y$ from Eq.~\eqref{eq:opt_X} and \eqref{eq:phase_opt_control} with parameters in Tab.~\ref{tab:pi_pulse_g}. The square pulses correspond to very short square pulses in $\omega_X$ and $\omega_Y$. Note that the square pulses are highly deformed by the response function of the cavity. They were used experimentally e.g. in~\cite{bienfait:2015} and they will be considered below as a reference of the control process. The spins are initially assumed to be along the $z$- axis with a polarization given by the Purcell effect and interacting with a cavity with zero photon. We neglect in the different numerical simulations the relaxation times $T_1$ and $T_2$. As could be expected, we observe in Fig.~\ref{figrobust} that the efficiency of bump fields is preserved for a wide range of offset frequencies, while a very good fidelity against variation of the $g$- parameter is achieved on resonance for the $g$- robust pulse. Bump and $g$- robust solutions lead to a more robust control protocol than the standard square pulses.
As a second point of comparison, we study the performance of $g$- robust, bump and square pulses in the maximization of the SNR. For each control sequence, we consider two cases, one corresponding to a constant $g=g_0$ distribution and the second to inhomogeneities of the form $g = g_0(1+\alpha)$, with $\alpha \in [-0.3,0.3]$. We also simulate ideal rotations on the spin system in order to estimate the maximum echo signal that can be reached with the spin distribution. The numerical results are displayed in Fig.~\ref{figecho}, which shows the echo signal observed with the different pulses. For sake of comparison, the duration of the bump pulses is the same as the one of $g$- robust fields (better results could be obtained with shorter bump pulses). An echo with a higher amplitude and a shorter time is achieved with the optimal solutions. We observe that the area of the different echos in Fig.~\ref{figecho} is roughly the same for the different excitations. However, due to the shorter echo, the SNR is indeed increased with the optimal pulses. 

\begin{figure}[h]
\includegraphics[width=7.5cm]{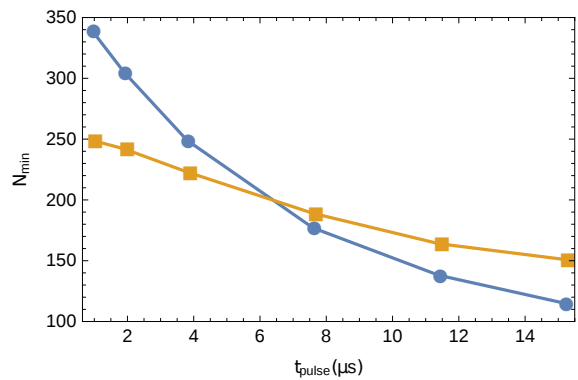}
\caption{(Color online) Number of spins $N_{min}$ per echo for bump (blue) and square (orange) $\omega_X$- pulses of duration $t_{\textrm{pulse}}$. The parameter $\kappa$ is set to $9.8\times 10^5$~s$^{-1}$.}
\label{fig6}
\end{figure}

The last point concerns the minimum number of spins $N_{min}$ per echo with a SNR of 1 which can be excited by the different control protocols. As can be seen in Fig.~\ref{fig6}, this minimum number is of the order of 260 spins with a square pulse of duration 1~$\mu$s~\cite{bienfait:2015,prost:2017}. In the case of Fig.~\ref{fig6}, we show that a smaller number of spins for a SNR of 1 can be excited with bump pulses. $N_{min}$ is of the order of 100 for a 15~$\mu$s long control process. This result is not obvious since bump pulses lead to higher SNR but at the price of a larger number of excited spins. In other words, the control time $t_k=1/k$ has to be adjusted to improve the selectivity of the excitation process without reducing drastically the SNR. In this case, $g$- robust $\pi/2$ pulses  are not good candidates for the minimization of  $N_{min}$ due to their strong robustness.

As a second example, we consider a CPMG sequence in which a series of $\pi$- pulses is applied periodically~\cite{levitt_book,coryCPMG:2010} with a period $T$ after the excitation process. We assume a perfect initial $\pi/2$- excitation of the spin ensemble and the different relaxation effects are accounted for. The total SNR after $M_r$ echo signals can be expressed as:
\begin{equation}
\textrm{SNR}_{M_r} = \sqrt{\sum_k^{M_r} \textrm{SNR}(k)^2},
\end{equation}
where $\textrm{SNR}(k)$ is the SNR of the $k$th echo.  The parameters of the different CPMG sequences are given in Tab.~\ref{tab:SNR_CPMG}. The time $T$ has been fixed to its minimum value for each pulse sequence. Table~\ref{tab:SNR_CPMG} gives the normalized SNR for one echo and for the whole CPMG sequence.

We observe in Tab.~\ref{tab:SNR_CPMG} that the $g$- robust pulses give a better SNR than the bump pulses. A noticeable enhancement of 25\% is also obtained with respect to the square pulses. The results achieved with the two optimal solutions are in addition very close to the physical limit of an ideal spin echo sequence. The improvement is even more striking for a CPMG sequence for which the gain is of the order of 60\%. Due to its short duration which allows a larger number of repetition, the bump pulse gives in this case the best result.

\begin{table}[h]
\centering{
\begin{tabular}{|c|c|c|c|c|c|}
\hline
\rule[-1ex]{0pt}{5ex} $\pi$- pulse & $\min (t_f)$ (s) & $M_r$ & $\frac{\textrm{SNR}_1}{\textrm{SNR}_{1}^{max}}$ & $\frac{\textrm{SNR}_{M_r}}{\textrm{SNR}_{Mr}^{max}}$ & $N_{\textrm{min}}$ \\
\hline
\hline
\rule[-1ex]{0pt}{5ex} Phys. limit  & $1.3 \times 10^{-5}$ & 321 & 1 & 1 & 55 \\
\hline
\rule[-1ex]{0pt}{5ex} Bump  & $3.69 \times 10^{-5}$ & 240 & 0.907 & 0.957 & 65  \\
\hline
\rule[-1ex]{0pt}{5ex} $g$- robust & $3.8 \times 10^{-5}$ & 231 & 0.967 & 0.918 & 67\\
\hline
\rule[-1ex]{0pt}{5ex} Square & $6.1 \times 10^{-5} $ & 160 & 0.754 & 0.406 & 94\\
\hline
\end{tabular} }
\caption{Parameters of the different CPMG sequences. The physical limit corresponds to ideal $\pi$- rotations for all the spins of the ensemble. $M_r$ indicates the maximum number of echos which can be observed. The number of spins $N_{\textrm{min}}$ for a SNR of 1 is computed for the first 100 echos of the sequence.}
\label{tab:SNR_CPMG}
\end{table}
\subsection{The high cooperativity regime}
\begin{figure*}[h]
\includegraphics[width=17.5cm]{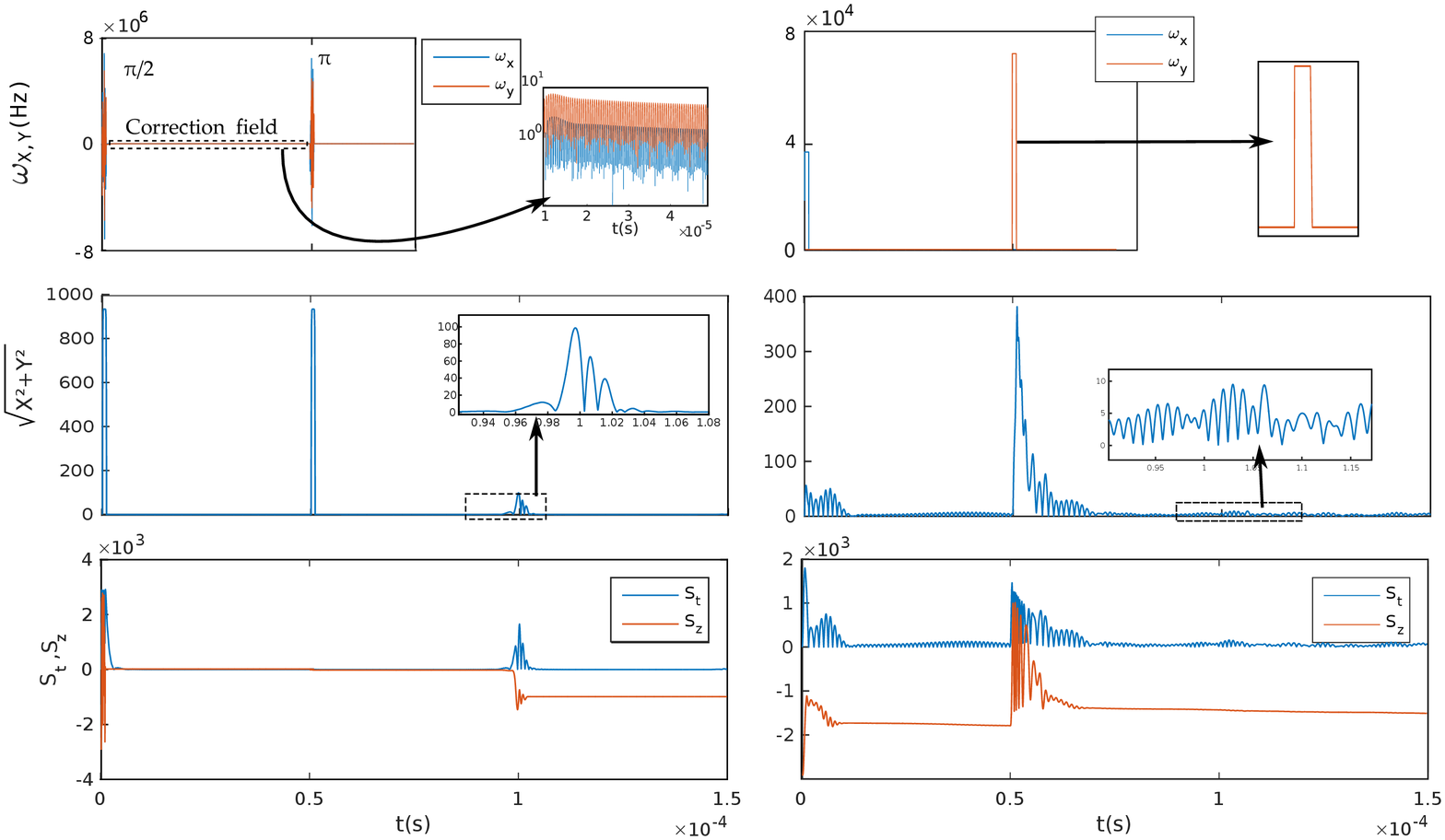}
\caption{(Color online) Spin echo sequence for $\Delta$- robust (left) and square pulses (right). The top panels display the control fields. Note the correction field applied between the $\pi/2$ and $\pi$ rotations in the first case. Time evolution of $\sqrt{X^2+Y^2}$ (Middle) and of the transverse $S_t = \sqrt{S_x^2+S_y^2}$ and longitudinal $S_z$ spin state.}
\label{fig7}
\end{figure*}

We investigate in this paragraph the controlled dynamics of a system in the high cooperativity regime ($C\gg 1$)~\cite{high1,high2}. For that purpose, we consider a new set of parameters: $N=135000$, $\gamma_r=100$~Hz and $g_0/(2\pi)=4240$~Hz. This leads to $N_{\textrm{eff}}=2916$, $\Omega_p=310$~kHz and $C=27$. The coupling strength $g$ is taken as a constant, $g=g_0$, and we neglect the relaxation times $T_1$ and $T_2$.  Figure~\ref{fig7} illustrates the dynamics induced by $\Delta$- robust and square pulses. The duration of each pulse is set to $1~\mu$s. Robust pulses are computed with the parameters presented in Tab.~\ref{tab:piHalf_pulse_Delta} and \ref{tab:pi_pulse_Delta}. Figure~\ref{fig7} clearly shows that the back action of the spin dynamics onto the control field given by Eq.~\eqref{eq:control} cannot be neglected. The field is different from zero during the whole control time and in particular between the $\pi/2$ and $\pi$- pulses.

An echo signal occurs in the cavity at $t=10^{-4}$~s. The shape of the echo is preserved with the optimal pulse, while a deformed echo with many small bumps is observed in the standard case. Furthermore, the intensity of the echo and the SNR are respectively increased by a factor 10 and 100. For a SNR of 1, we obtain $N_{\textrm{min}}=43$ and $N_{\textrm{min}}=53$ for the optimal and square pulses. The same observation can be made on the spin dynamics in Fig.~\ref{fig7}. In contrast to the variations produced by the square pulses, a perfect excitation can be seen with the optimal control, $S_z\simeq 0$ for $0<t<10^{-4}$. A strong relaxation due to the radiation damping occurs during the two echos. This signature of the high cooperativity regime can be observed when $\bar{S}_x$ and $\bar{S}_y$ are very large.

\section{Conclusion}\label{sec6}
In this work, we have applied quantum control techniques to an inhomogeneous spin ensemble coupled to a cavity. We have described a general optimization procedure allowing to implement standard Hahn and CPMG sequences in presence of offset and coupling strength inhomogeneities. Relevant experimental values in ESR have been used for the numerical simulations. Different control strategies extending from bump pulses to $g$- robust fields have been derived. The relative advantages of the different solutions have been discussed in the low and high cooperativity regimes. Their superiority in terms of SNR and sensitivity over the standard square pulses has been demonstrated. In the experimental setup under study, the numerical results show that a good compromise is provided by the bump pulses which combine simplicity, efficiency and robustness against offset inhomogeneities. Their short duration is also a key factor for the enhancement of the SNR by CPMG sequences.

These results confirm the key role that OCT could play in a near future for the detection of a single spin in ESR. Finally, we point out that more general pulse sequences could also be used to improve the efficiency of the overall process. An example is given by the cooperative pulses, a set of individual pulses which are designed to compensate each other's imperfections~\cite{coop2010,coop2018}. Recent works have shown their efficiency in a standard Hahn echo sequence. As a further step, it would be interesting to estimate the performance of such pulses in the maximization of the SNR or the minimization of spins per echo.\\ \\
\noindent\textbf{ACKNOWLEDGMENT}\\
S.J. Glaser acknowledges support from the DFG (Gl 203/7-2). D. Sugny and S. J. Glaser acknowledge support from the ANR-DFG research program Explosys (ANR-14-CE35-0013-01). D. Sugny acknowledges support from the PICS program and from the ANR-DFG research program COQS (ANR-15-CE30-0023-01). The work of D. Sugny has been done with the support of the Technische Universit\"at M\"unchen – Institute for Advanced Study, funded by the German Excellence Initiative and the European Union Seventh Framework Programme under grant agreement 291763. This project has received funding from the European Union's Horizon 2020 research and innovation programme under the Marie-Sklodowska-Curie grant
agreement No 765267.

\appendix

\section{parameters of optimal pulses}\label{appa}
This paragraph provides the different set of parameters for the optimal pulses used in this work.
\begin{table}[h]
\centering
\small{
\begin{tabular}{@{}|c|c|c|@{}}
\toprule
$p$ & \multicolumn{2}{c|}{10} \\
$t_f$ & \multicolumn{2}{c|}{1} \\
$\mathcal{F}$ & \multicolumn{2}{c|}{0.9993}\\
\toprule
$n$ & $a_n$ & $b_n$ \\
\toprule
0 & 2.48502852519278   &	0 \\
1 &-0.614602837937966  &	0.0222236529656774 \\
2 &-0.146403432037310  &	-0.326319502810118 \\
3 &0.249569148521250   &	0.212035090021068 \\
4 &-0.380514318815982  &	-0.294315446425150 \\
5 &-0.850981648334035  &	0.292006472615227 \\
6 &0.00534202375558939 &	-0.284521506361719 \\
7 &-0.445742825754110  &	-0.00924269034857846 \\
\toprule
\end{tabular}}
\caption{List of parameters for a $\Delta$- robust $(\pi/2)_y$ universal rotation.}
\label{tab:piHalf_pulse_Delta}
\end{table}

\begin{table}[h]
\centering
\begin{tabular}{@{}|c|c|c|@{}}
\toprule
$p$ & \multicolumn{2}{c|}{10} \\
$t_f$ & \multicolumn{2}{c|}{1} \\
$\mathcal{F}$ & \multicolumn{2}{c|}{0.9997}\\
\toprule
$n$ & $a_n$ & $b_n$ \\
\toprule
0 & 3.6552961005  &	 0 \\
1 & -0.1862900729 &	 0.3016414772 \\
2 & 0.1569621446  &	 0.9517460473 \\
3 & 0.886144687	  &  -0.5237345127 \\
4 & -0.3948819182 &	 0.439535574 \\
5 & -0.362518991  &	 -0.261853447 \\
6 & 0.1747816746  &	 0.2785205439 \\
7 & 0.0332864557  & 	0.0097613015 \\
\toprule
\end{tabular}
\caption{List of parameters for a $\Delta$- robust $(\pi)_y$ universal rotation}
\label{tab:pi_pulse_Delta}
\end{table}
\begin{table}[h]
\centering
\small{
\begin{tabular}{@{}|c|c|@{}}
\toprule
$p$ & \multicolumn{1}{c|}{2}\\
$t_f$ & \multicolumn{1}{c|}{1}\\
$\mathcal{F}$ & \multicolumn{1}{c|}{0.999}\\
\toprule
$n$ & $a_n$ \\
\toprule
0 & 1.45730821080502   \\
1 & -1.90458549438015  \\
2 & 0.471852675517646  \\
3 & -0.164591020002767 \\
4 & 0.691022251640240  \\
\toprule
\end{tabular}}
\small{
\begin{tabular}{@{}|c|c|@{}}
\toprule
$p$ & \multicolumn{1}{c|}{2} \\
$t_f$ & \multicolumn{1}{c|}{1} \\
$\mathcal{F}$ & \multicolumn{1}{c|}{0.999}\\
\toprule
$n$ & $a_n$ \\
\toprule
0 & 1,05923686097438   \\
1 & -1,06434468127802  \\
2 & 0,197782131275470  \\
3 & -0,985850874873962 \\
4 & -0,680625181005274 \\
5 & -0,680625181005274 \\
\toprule
\end{tabular}}
\caption{List of parameters for $g$- robust $(\pi/2)_y$ (left) and $(\pi)_y$ (right) universal rotations. The $b_n$ parameters are fixed to 0.}
\label{tab:pi_pulse_g}
\end{table}

\end{document}